# Electric Field Induced Transformation of Magnetic Domain Structure in Magnetoelectric Iron Garnet Films

*A.P. Pyatakov[*], G.A. Meshkov, A.V. Nikolaev, E.P. Nikolaeva, A.S. Logginov,*

*Physics Department, M.V. Lomonosov MSU, Leninskie gori, Moscow, 119992, Russia;*

*and A.K. Zvezdin*

*A.M. Prokhorov General Physics Institute, 38, Vavilova st. Moscow, 119991, Russia;*

\* Corresponding author: pyatakov@phys.msu.ru

The room temperature magnetoelectric effect was observed in epitaxial iron garnet films that appeared as magnetic domain wall motion induced by electric field. The films grown on gadolinium-gallium garnet substrates with various crystallographic orientations were examined. The effect was observed in (210) and (110) films and was not observed in (111) films. Dynamic observation of the domain wall motion in 400 V voltage pulses gave the value of domain wall velocity in the range 30÷50 m/s. The same velocity was achieved in magnetic field pulse about 50 Oe.

**Introduction**

During last few years the revival of the interest to magnetoelectric media is observed [1-5]. In this type of materials there is a coupling between magnetic and electric properties that provide a novel approach to the magnetic/electric field conversion that can break through the barriers for increasing storage density in magnetic memory and miniaturization of spin electronic devices that suffer from energy losses and progressive damage of metal conductors caused by electric currents. The considerable progress in the area of magnetoelectric materials has been achieved, especially in thin-film deposition techniques [6]. There were reports on various electric field control of magnetism [7-12], but all of them have at least one of the following limitations:

- magnetoelectric properties appear at low temperatures, as in the majority of multiferroic materials [7-9]
- among not numerous room temperature magnetoelectrics the most of them are antiferromagnetic materials, thus their magnetic properties are compensated. Special exchange coupled antiferromagnetic/ferromagnetic structure is needed to convert electrical switching of antiferromagnetic order parameter into magnetization switching [10].
- In artificial magnetoelectric materials (composites) the electric and magnetic subsystems are spatially separated [11, 12]. The dependence of the coupling between two subsystems on the conditions at the interfaces and their local variations results in unpredictable character of magnetization switching [11] and intricate magnetic stripe-domain pattern [12].

In paper [13] the effect of electric field driven magnetic domain wall motion in iron garnet films was discovered. This sort of magnetoelectric effect provides with the control of magnetization that is realized in single phase material at room temperature by usage electric field only, not implying magnetic field or charge carriers transport. In this paper we report on the results of static and dynamic measurements of the magnetoelectric control of domain walls in films grown on gadolinium-gallium garnet substrates with various crystallographic orientation. The simple theoretical model to explain the basic features of the effect is provided.

In our experiments we used iron garnet films grown by liquid-phase epitaxy on (111), (110), and (210) $Gd_3Ga_5O_{12}$ substrate. The relevant parameters of the samples are listed in Table 1. To produce a high-strength electric field in the dielectric iron garnet film, we used a 50μm-diameter copper wire with a pointed tip, which touched the surface of the sample in the vicinity of the domain wall (Fig. 1 and inset). The tip curvature radius of the copper "needle" was about 5 μm. This allowed us to obtain an electric field strength of up to 1000 kV/cm near the tip by supplying a voltage of up to 500 V to the needle. The field caused no dielectric breakdown, because it decreased rapidly with distance from the tip and, near the grounding electrode (a metal foil attached to the substrate), did not exceed 500 V/cm. The absence of the possible leakage currents between the tip and the grounding electrode (e.g., over the sample surface) was verified by a milliamperemeter. The magnetooptical technique in Faraday geometry was used to observe the micromagnetic structure through a hole ~ 0.3 mm in diameter that was made in the grounding electrode. The image of the magnetic structure was taken by CCD camera connected with a personal computer.

For dynamic measurements the high speed photography technique was used: the pulses of electric field (pulse width ~ 300ns, the rise time ~20ns) were followed by pulses of laser illumination (duration ~10ns) to get an instantaneous image of the structure under the influence of electric field. Varying the time delay between field and laser pulses enabled us to observe the consecutive positions of domain wall and thus investigate its dynamics. In the dynamic measurements the amplitude of voltage pulses was up to 400V.

In static measurements we register the magnetization distribution in initial state, then after electric field was switched on and finally at electric field switched off. As a result, we obtained series of images for different voltage polarities (Fig. 2). We observed a local displacement of the domain walls in the vicinity of the tip. This effect of electric field controlled domain wall position was observed in iron garnets films with (210) and (110) substrate orientation and was not observed in (111) films (Table 1, note that to illustrate idea the Table 1 lists the most representative examples while much more samples were tested to verify the dependence on substrate orientation). The magnitude of the displacement increased with voltage. The most prominent changes were observed in (210) films at stripe domain heads and bubble domains (Fig. 2). As soon as the dc voltage was switched off the domain walls came back to the equilibrium positions (fig 2 a). Reversing the polarity of the voltage caused the opposite changes in micromagnetic structure (compare Fig. 2b and 2c). The reversible domain wall displacements up to 5μm were observed. At higher values of displacement the modification of the micromagnetic structure had irreversible character, e.g. the percolation of the bubble domain with the nearest stripe domain head (Fig. 2d).

The results of observation in pulsed electric field show that in response to the applied electric field domain wall steadily moves towards its new equilibrium position. Depending on the amplitude of the electric field pulse and the distance between the needle tip and the wall, the wall moves within 10 – 100 ns after the pulse start.

To compare the velocities achieved in electric field with typical velocities of domain wall in magnetic field we carried out the measurements in magnetic field pulses. The velocity of 50m/s similar to that one obtained in voltage pulse 400V (electric field E=800kV/cm) was achieved in pulse magnetic field about 50 Oe.

We point out several characteristic features of the phenomenon, which serve as the basis for the following discussion:

(i) The direction of the domain wall displacement depends on the polarity of the voltage (and, hence, on the direction of the electric field): in the case of positive polarity, the wall was attracted to the needle, and, in the case of negative polarity, it was repulsed.

(ii) The direction of the wall displacement did not depend on the direction of magnetization in the domain (along the $z$ axis or opposite to it, see the inset in Fig. 1).

(iii) The effect was observed in films with considerable in-plane anisotropy ((210) and (110) substrate orientations) and was not observed in highly symmetrical (111) films.

The characteristic features listed above allow us to exclude the effects of non-magnetoelectric nature that could lead to displacements of domain walls: the magnetic fields caused by possible leakage currents and the magnetostrictive phenomena caused by the pressure of the tip on the sample due to electrostatic attraction. Indeed, the dependence on the polarity of the voltage applied to the needle (feature (i)) allows us to exclude the effect of the tip pressure on the sample, because the tip polarizes the sample surface and is attracted to it irrespective of the sign of the potential at the needle; hence, the effect caused by the tip pressure should be independent of polarity. Feature (ii) testifies that effect cannot be related to the parasitic magnetic moment of the tip or magnetic field of leakage currents because, otherwise, the domain walls would be displaced in opposite directions for domains with opposite magnetization. Thus, features (i) and (ii) of the phenomenon under study allow us to conclude that the latter is of magnetoelectric nature.

The feature (iii) highlights the role of the crystallographic symmetry of the films. In bulk iron garnet samples, due to the presence of an inversion center in the crystal symmetry group, only the effects proportional to even powers of electric field, represented by quadratic magnetoelectric [14] and electromagnetooptical [15] effects, are possible. The dependence of the direction of the domain wall displacement on the electric polarity (the oddness of the effect with respect to electric field) testifies to the violation of the space inversion in films, unlike the case of bulk iron garnet samples. This conclusion is supported by reports on observation in iron garnet films the linear electro-magnetooptical effect [16] and second harmonic optical generation [17,18], that are allowed only in media with broken space-inversion symmetry. Moreover, the effect of domain wall displacement was the most strongly pronounced in low symmetry (210) films (point group symmetry *m*), while highly symmetrical (111) films (point group symmetry *3m*) demonstrate bulk-like behavior (Table 1, the right column). It is interesting to note that in analogy to our effect the linear electro-magnetooptical effect [16] and second harmonic generation [18] were also pronounced in (210) films (one or two orders of magnitude larger as compared to other film orientations).

The influence of electric field on micromagnetic structure was predicted theoretically in the series of works [19-23]. These theoretical models took into account the so-called *inhomogeneous magnetoelectric interaction* that gives rise to electric polarization associated with magnetic inhomogeneities. From this point of view such spatially modulated magnetic structures as magnetic domain walls [19,21], spin cycloid [20], magnetic vortices [22] and vertical Bloch line [23] were considered and it was shown that various electric charge distribution are associated with them.

The inhomogeneous magnetoelectric effect corresponds to the following contribution to the free energy of the crystal:

$$F_{ME} = \gamma_{ijkl} \cdot E_i \cdot M_j \cdot \nabla_k M_l, \qquad (1)$$

where $\mathbf{M}=\mathbf{M}(\mathbf{r})$ is magnetization distribution, $\mathbf{E}$ is electric field, $\nabla$ is vector differential operator, $\gamma_{ijkl}$ is the tensor of inhomogeneous magnetoelectric that is determined by the symmetry of the crystal. One can learn immediately from the equation (1) that the effect is odd in electric field $\mathbf{E}$, and doesn't change the sign with magnetization $\mathbf{M}$ reversal, that agrees with the features (i) and (ii) of the effect.

The inhomogeneous magnetoelectric contributions (1) for the bulk crystal of iron garnets with cubic symmetry takes the following high symmetry form [20, 22]:

$$F_{ME} = \gamma \cdot \mathbf{E} \cdot \left( \mathbf{M} \cdot (\nabla \cdot \mathbf{M}) - (\mathbf{M} \cdot \nabla)\mathbf{M} \right) \qquad (2)$$

The feature (iii) of the effect, i.e. the dependence on the substrate orientation, can be explained by the fact that the necessary condition for the effect is the local violation of space inversion in domain wall, i.e. the nonzero $(\nabla \cdot \mathbf{M})$ and $(\mathbf{M} \cdot \nabla)\mathbf{M}$ terms. In (111) films we deal with the Bloch type domain walls ($(\nabla \cdot \mathbf{M}) = 0$, $(\mathbf{M} \cdot \nabla)\mathbf{M} = 0$) the space inversion symmetry characteristic for the bulk material persists and thus electric field have no effect on domain wall. However the anisotropy of (210) and (110) samples requires the deflection of the magnetization from the normal to the film resulting in Neel component in the domain wall: the direction of spin modulation and normal to the rotation plane do not coincide any more. This condition is expressed in mathematical form as $(\nabla \cdot \mathbf{M}) \neq 0$, $(\mathbf{M} \cdot \nabla)\mathbf{M} \neq 0$, and thus nonzero magnetoelectric terms in (2).

Considering the effect of electric field on micromagnetic structure in iron garnet films we should also mention another possible mechanism that was involved to explain the electro-magnetooptical effect in iron garnets films [24] governed by changes of magnetic anisotropy induced by electric field.

In conclusion, the electric field control of magnetization distribution is implemented in single crystal material at room temperature not implying electric current. The effect is observed in epitaxial iron garnet films with in-plane anisotropy (grown on (210) and (110) gadolinium-gallium garnet substrates) and is not observed in high symmetry (111) films. The direction of domain wall displacement under the influence of electric field is dependent on the electric polarity and independent of the direction of magnetization in the domains. The domain wall displacement has reversible character in the range of 1÷5 μm and irreversible one at larger distances. The average domain wall velocity of 50 m/s in 800kV/cm electric field pulses was achieved that was equivalent to the effect of 50 Oe magnetic field pulses. Although the most part of the measurements was done at voltages of about 500V the effect of electric field was still discernible at voltages of 100V and this value can be scaled down to several Volts by further miniaturization of the electrode to nanometric size (the curvature radius of one used in experiment was ~5μm). This effect opens new exciting possibilities in the field of micro- and nanomagnetism providing the means for electric field control of magnetization distribution.

**Table 1**

Parameters of the samples under study and magnetoelectric control effect registration marks. Symbols h stands for thickness of the iron garnet film, $M_S$ is saturation magnetization, p is a period of domain structure. At the right column the presence/abscence of the magnetic domain wall displacement in electric field is indicated.

| N | substrate orientation | Chem. composition | h, μm | $4\pi M_s$, G | p, μm | Effect detection |
|---|---|---|---|---|---|---|
| 1 | (111) | $(BiTm)_3(FeGa)_5O_{12}$ | 10 | 144 | 8,7 | no |
| 2 | (111) | $(BiLu)_3(FeGa)_5O_{12}$ | 19 | 78 | 39 | no |
| 3 | (110) | $(BiLu)_3(FeGa)_5O_{12}$ | 4 | 162 | 9,2 | yes |
| 4 | (110) | $(BiLu)_3(FeGa)_5O_{12}$ | 6 | 76 | 14,4 | yes |
| 5 | (210) | $(BiLu)_3(FeGa)_5O_{12}$ | 10 | 53.5 | 34 | yes |
| 6 | (210) | $(BiLu)_3(FeGa)_5O_{12}$ | 10 | 62 | 28 | yes |
| 7 | (210) | $(BiLu)_3(FeGa)_5O_{12}$ | 8.6 | 55 | 27 | yes |

**Figure 1** Schematic representation of the geometry of the experiment and the configurations of the electric field and magnetization. The electric field (the field lines are shown by the dashed lines) is formed in the dielectric medium of the sample between the needle (*1*) and the metal foil (*2*), which plays the role of the grounding electrode. The maximum field strength (about 1 MV/cm) is reached in the magnetic film (*3*) near the tip; it decreases rapidly in the bulk of the substrate (*4*) and does not exceed 500 V/cm near the grounding electrode (*2*). The abscence of the leakage currents is controled with the milliampermeter (mA). The incident light (denoted with wavy arrows) is along the normal to the surface. The objective lens (5) is placed behind the pinhole in the foil (2). The inset shows the magnetization distribution in the film: the domain wall (*6*) separates two domains (*7, 8*) with opposite magnetization directions; the tip (*9*) touches the iron garnet surface near the domain wall.

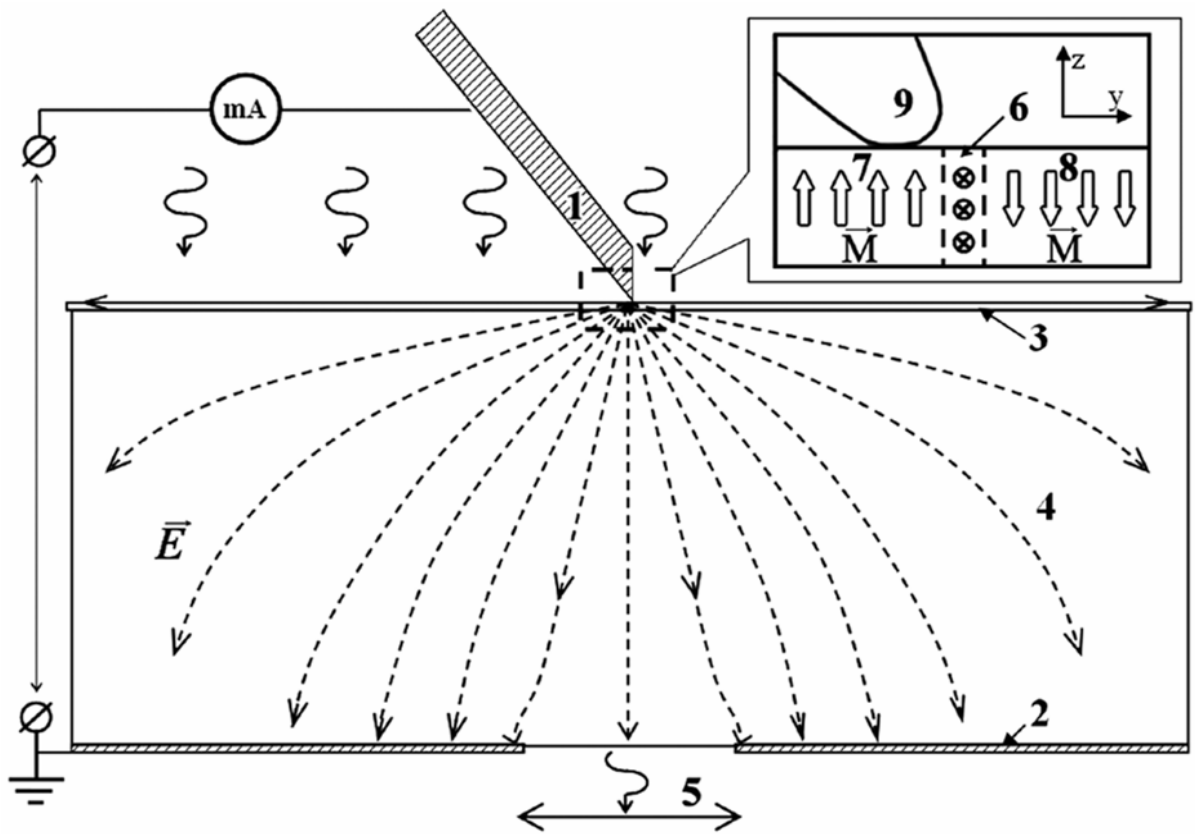

**Figure 2** Domain walls displacement under the influence of static electric field applied. a) Initial state with no voltage applied: (1) is the tip, (2) is bubble domain, (3) is domain head. b) displacement of the stripe domain head and bubble domain nearest to the tip-sample contact towards the tip at electric potential +500V at the tip, c) the opposite displacement at negative potential -500 V at the tip, d) the irreversible changes of micromagnetic structure after application of higher voltage +1500V to the tip. The measurements were done with the sample 6, Table 1 (the blurring of the images of stripe-domain heads is characteristic feature of the (210) films caused by the deflection of magnetic easy axis from the normal to the film).

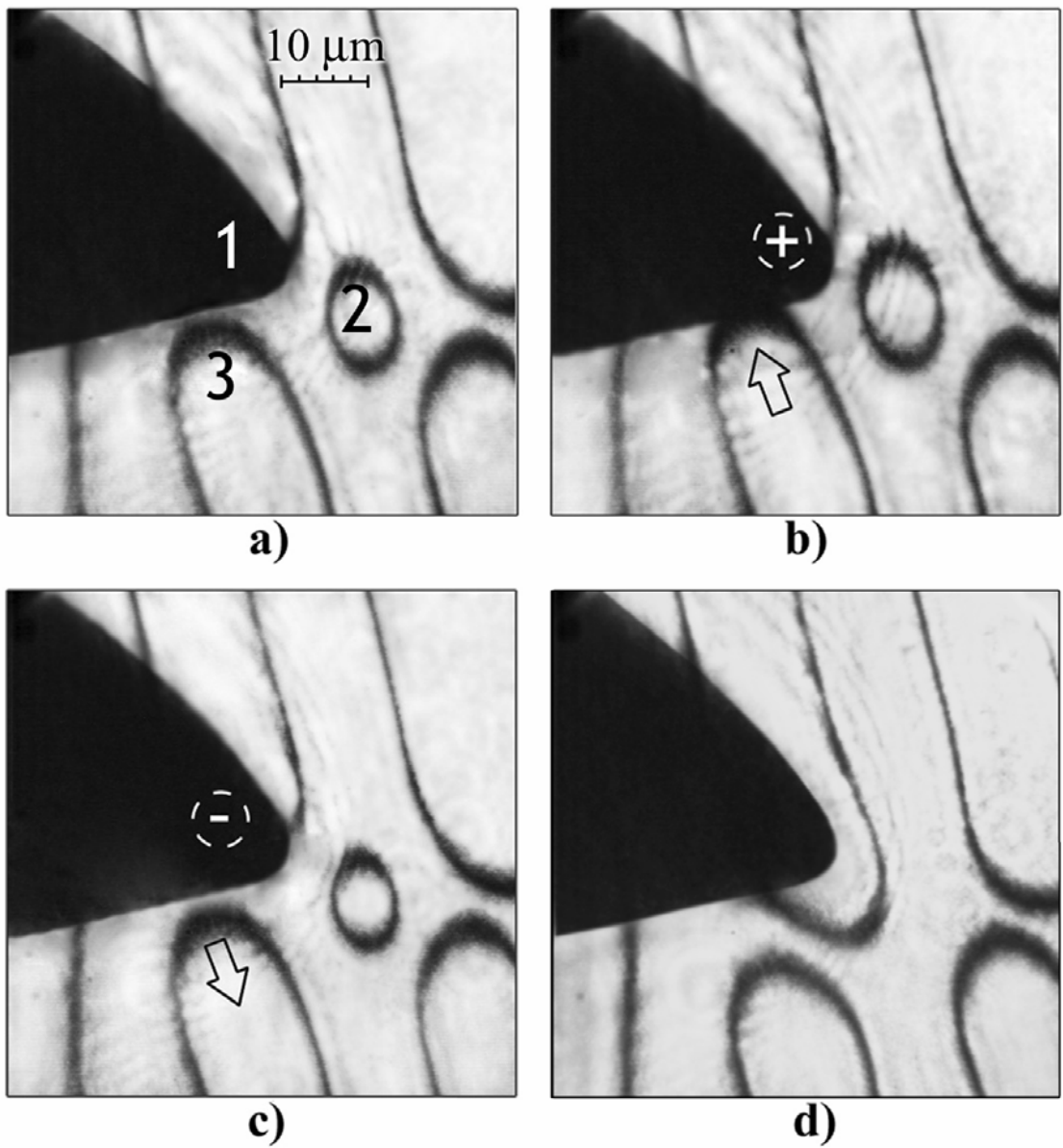